# PLANETARY NEBULAE AS PROBES OF DARK MATTER IN NGC 3384

Benoit Tremblay, David Merritt and T. B. Williams
Department of Physics and Astronomy, Rutgers University,
Piscataway, NJ 08855

ABSTRACT

We have obtained radial velocities of 68 planetary nebulae (PNs) surrounding the SB0 galaxy NGC 3384 in the Leo I group, using the CTIO 4 m telescope and the Rutgers Fabry-Perot interferometer. The PN system exhibits a well-ordered rotation field aligned with the photometric axes of the galaxy. The rotation curve is flat from about 2 kpc until at least 7 kpc. We infer a lower limit of $0.7 \times 10^{11} M_\odot$ for the mass of NGC 3384. Our results imply that at least a third of the dynamical mass of the NGC 3379/3384 system – estimated by Schneider (1985) to be roughly $6 \times 10^{11} M_\odot$ – may be accounted for in the two bright galaxies.

## 1. INTRODUCTION

Planetary nebulae (PNs) are bright emission line objects. Their usefulness as extragalactic distance indicators has been appreciated for some time (Jacoby 1989; Ciardullo *et al.* 1989), but it is only recently that Hui (1993), Ciardullo *et al.* (1993), and Arnaboli *et al.* (1994) have demonstrated their utility as kinematic tracers. PNs are well suited to probing the outer regions of early type galaxies for several reasons. They are distributed spatially like the underlying galactic light, so that significant numbers of PNs will be found at large radii. Their luminosities are uncorrelated with position in the galaxy, so they are measurable even in the outer parts of the galaxy. Their flux is concentrated in a few strong emission lines, greatly enhancing their contrast against the underlying galaxy and providing an immediate velocity determination.

We are currently using an imaging Fabry-Perot (FP) interferometer to investigate the kinematics of PN systems surrounding early type galaxies. This approach provides a powerful means for simultaneously detecting and measuring the velocities of all the PN in a relatively large field of view about the galaxy. By using a FP etalon with spectral resolution slightly broader than the width of the PN line, we maximize the contrast of the PN against the background light of the galaxy and the night sky. We can thus detect PNs that are far from the peak of the luminosity function, and obtain large samples of discrete velocities for mapping the gravitational potential of the galaxy.

While FP scans provide only a short portion of the spectrum of each object, this is entirely adequate for PNs, where all of the flux is in a few emission lines scattered over the spectrum. For most PNs, the [OIII] 5007Å line contains half or more of the total flux, so little is lost by concentrating our observations in this spectral region, while the multiplex gains of simultaneously measuring tens to hundreds of PN spectra are significant. The line profiles from the FP observations yield mean velocities with precisions that exceed the demands of galactic mass determinations. The Rutgers Fabry-Perot (RFP) system has significantly higher throughput than most spectrographic instruments (see Schommer *et al.* 1993), so this observing technique is very efficient. Finally, since the PNs are detected and have their velocities measured from *images* of the galaxy, no precise positioning of fibers or slitlets is required, greatly simplifying the observations.

Here we report the first results from a FP study of PNs in galaxies of the Leo I group. We have obtained velocities of 68 PNs in the vicinity of NGC 3384. Our sample contains almost all of the PNs previously identified by Ciardullo *et al.* that lie within three, 2.7 arcmin diameter fields centered on and offset from NGC 3384. We find that the PNs exhibit a well-ordered rotation field, with a flat rotation curve extending to the edge of the sample, roughly 7 kpc from the center of NGC 3384. We derive a lower limit on the mass of the galaxy and compare it to the masses derived by Ciardullo *et al.* for the companion E0 galaxy NGC 3379, and to the total mass of the 3379/3384 system as derived by Schneider (1985) from the kinematics of the surrounding HI ring. We are thus able to begin to assess how the total mass of the Leo I group is partitioned among its bright galaxies.

## 2. OBSERVATIONS AND DATA REDUCTION

We observed NGC 3384 on the nights of April 7 to April 10, 1994 with the CTIO 4m telescope and the RFP. We obtained a series of 2Å FWHM images about the [OIII] 5007Å emission line; the images were separated in wavelength by 1.25Å, with 19 images per field, covering a total velocity range of 1350 km s$^{-1}$. The exposure time for each image was 15 minutes. Filters of 49Å width, centered at 5007Å for etalon settings from 5011.75Å to 5018.0Å and at 5037Å for settings from 5019.25Å to 5036.75Å, were used to select the desired interference order.

The data presented here are from three fields obtained on four nights of observation. The field of the first night was offset 1.7 arc minutes to the west of the galaxy center, keeping the galaxy bulge in the eastern edge of the frame. The field of the second night was centered on the galaxy, and the field of the third night was offset 1 arc minute to the east. On the fourth night we repeated the field centered on the galaxy.

The images were flat–fielded, bias–subtracted, and trimmed in the normal way. Charged particle events were removed, and the images were shifted to a common reference position. The images were then combined to form a smooth template, which was subtracted from the individual images (through a linear least-squares fit) to remove the strong background gradient due to the galaxy. The presence of reference stars in each frame allowed the fluxes of the individual frames to be normalized using the DAOGROW prescription (Stetson 1990). These stars also provided us with a means to measure the point spread function (psf) characteristic of each frame. We used the DAOPHOT package (Stetson 1987) to derive the stellar psf in the unsubtracted frames.

PNs may be identified by searching in our position-wavelength data cube for an object with an emission-line spectrum and with the proper spatial psf. In this *Letter*, we present velocities of only those PNs whose positions were previously reported by Ciardullo *et al.*; in a later paper we will report positions and velocities for a much larger sample of PNs. Ciardullo *et al.* identified 102 PNs in the vicinity of NGC 3384, of which 71 lie within our fields. While there was a generally good correlation between our fluxes and those quoted by Ciardullo *et al.*, we were unable to find the PNs numbered 61, 63 and 81 in the Ciardullo *et al.* list even though these objects lay within at least one of our three fields.

Figure 1 shows spectra of three typical PNs from Ciardullo *et al.*'s list. Our estimated velocity uncertainty is typically $\sim 7$ km s$^{-1}$ for PNs present in more than one field (55 PN) and $\sim 30$ km s$^{-1}$ for those found in a single field (13 PN). Aside from the three missing objects mentioned above, we obtained good fits to the RFP instrumental profile for even the faintest PNs in the Ciardullo *et al.* list, which suggests that a complete analysis of our RFP data should yield a few hundred PN velocities with reasonable precision.

## 3. DYNAMICS OF THE NGC 3384 PLANETARY NEBULA SYSTEM

Figure 2 shows the line-of-sight mean velocity and velocity dispersion fields defined by the observed sample of 68 PNs. A systemic velocity of 735 km s$^{-1}$ was assumed (de Vaucouleurs *et* *al.* 1991). The rotational velocity field (Figure 2a) was computed from the measured velocities using a two-dimensional thin-plate smoothing spline (Wahba 1990, p. 30), as implemented in the GCVPACK Fortran library (Bates *et al.* 1986). This algorithm yields a smooth estimate of any function given a set of measured values with errors. The velocity dispersion field (Figure 2b) was computed in the same way by fitting a smoothing spline to the squared velocities and subtracting the mean velocity and the estimated velocity errors in quadrature.

The rotational velocity field of the PN system is reasonably symmetric, with the greatest rotation occurring along an axis that is approximately aligned with the major axis of NGC 3384. As Figure 2c shows, the line-of-sight mean velocity rises to a peak of about 125 km s$^{-1}$ at $\sim$2 kpc from the center of NGC 3384, then remains approximately constant out to the limit of the data, at $\sim$7 kpc. The isovelocity contours appear consistent with those of an inclined disk, suggesting that most or all of the PNs lie within a plane. However the observed sample is too small to verify this hypothesis. Assuming a disk geometry and an inclination of NGC 3384 of 67 degrees from face-on, the streaming velocity $\bar{v}_\phi$ (not the circular velocity) associated with the flat part of the PN rotation curve is about 140 km s$^{-1}$. This value is likely to be a slight underestimate because of the smoothing effect of the spline.

The velocity dispersion field (Figs. 2b and 2d) shows less apparent symmetry, although there is a hint of reflection symmetry about the principal axes of NGC 3384. The line-of-sight velocity dispersion defined by the PNs nearest to NGC 3384 is roughly 100 km s$^{-1}$; this may be compared to a value of 156 km s$^{-1}$ obtained by Ore *et al.* (1991) for the stellar core of NGC 3384. At larger radii, the dependence of the PN dispersion on radius is unclear from this modest sample, although it appears to drop in directions parallel to the galaxy major axis and to remain roughly constant along the minor axis.

The kinematics of the NGC 3384 PN system are dominated by rotation. The ratio $v/\sigma_0$ of the peak rotation velocity to the central velocity dispersion is about 1.4, and remains close to one even if the larger, stellar velocity dispersion is used as an estimate of $\sigma_0$. Furthermore, the number of PNs with retrograde velocities (*i.e.* PNs with line-of-sight velocities opposite in sign to that of the mean line-of-sight velocity at the same position) is small, amounting to about four out of 68 (Figure 2).

Estimating the three-dimensional form of the gravitational potential around NGC 3384 in a model-independent way is impossible from a modest kinematical sample like this one. We therefore adopt three simplifying assumptions. First, we assume that the PNs lie approximately in a circular disk that is coplanar with the disk of NGC 3384. This assumption is consistent with the shapes of the isovelocity contours and with the large value of $v/\sigma_0$; furthermore we expect the PNs to have roughly the same spatial distribution as the other stars in NGC 3384. Second, we assume that the velocity dispersion tensor defined by the PNs is isotropic about the mean rotation at every point – *i.e.* that the PN system constitutes an "isotropic oblate rotator." This assumption can not be verified from our data but seems reasonable as a first approximation. The assumption of a planar geometry implies that the true rotation velocity is related to the observed value by the sine of the inclination. The additional assumption of velocity isotropy implies that the intrinsic dispersions are equal to their observed values at every point, without no corrections required for inclination. Finally, we assume that most of the mass surrounding NGC 3384 is distributed in a spherical halo.

Our first two assumptions are mildly inconsistent, since an isotropic oblate rotator has a flattening that is determined by the ratio $v/\sigma_0$, and we have assumed that the PNs lie in an infinitely thin disk. However the inconsistency affects primarily the interpretation of the dispersions, and we show below that the random motions make only a minor contribution to the inferred mass.

Following our simplifying assumptions, the axisymmetric Jeans equation in the equatorial plane then yields for the mass $M$ within radius $r$:

$$GM(<r) = r\overline{v}_\phi^2(r) - r\sigma^2(r)\left(\frac{\partial \log \nu}{\partial \log r} + \frac{\partial \log \sigma^2}{\partial \log r}\right), \qquad (1)$$

with $\overline{v}_\phi(r)$ the mean rotational velocity defined by the PNs, $\sigma(r)$ their velocity dispersion, and $\nu(r)$ the PN number density profile. For $\sigma(r)$ we adopt the constant value of 75 km s$^{-1}$, since the radial dependence of the velocity dispersion is unclear from our data; in so doing we probably underestimate the mass at every radius since we ignore the contribution of the last term in equation (1). For $\nu(r)$ we assume the same dependence on radius as for the starlight in NGC 3384, which, outside of about 1 kpc, follows an exponential profile with a scale length of $\sim 4.5$ kpc.

We find that the largest contribution to the inferred mass outside of a few kpc comes from the $r\overline{v}_\phi^2(r)$ term in Equation (1). $M(<r)$ increases roughly linearly with radius between 2 and 7 kpc, giving a total mass of $\sim 0.65 \times 10^{11} M_\odot$ at $r = 7$ kpc.

The apparent blue magnitude of NGC 3384 is 10.75 (de Vaucouleurs *et al.* 1991), corresponding to an absolute blue magnitude of -19.27 at our assumed distance of 10.1 Mpc (Ciardullo *et al.* 1989). Combined with our estimate for the mass within 7 kpc, the blue mass-to-light ratio for NGC 3384 is roughly 9 in solar units. This figure should almost certainly be viewed as a lower limit since both the mean and random velocities of the PNs remain high at the limits of our sample.

### 4. THE DISTRIBUTION OF MASS IN THE LEO I GROUP

NGC 3384 and its E0 companion NGC 3379 are surrounded by a 200 kpc diameter HI ring (Schneider *et al.* 1989). From measurements of the variation of radial velocity around the ring, Schneider (1985, 1991) solved for the parameters of the Keplerian orbit described by the gas and inferred an enclosed mass of $5.6 \times 10^{11} M_\odot$. One focus of the elliptical orbit appears to lie close to the 3379/3384 pair. From the persistence of the ring as an ellipse, Schneider (1991) concluded that the force law is roughly inverse square and therefore that most of the mass in the system lies in the near vicinity of the two bright galaxies, no more than $\sim 60$ kpc from the center of either one.

Ciardullo *et al.* (1993) used radial velocities of a sample of 29 PNs to estimate a mass of $\sim 1.2 \times 10^{11} M_\odot$ for NGC 3379. The dynamics of the NGC 3379 PN system are dominated by random motions, and the Ciardullo *et al.* mass estimate is correspondingly very uncertain; it depends sensitively on their assumption that mass follows light. A mass distribution that is more extended than the stellar light in NGC 3379 would require a larger total mass to reproduce their kinematical data, and could do so equally well (e.g. Dejonghe & Merritt 1992). (Our estimate of the mass of NGC 3384 is considerably less uncertain because the 3384 PN system is strongly rotating, making the unknown shape of the velocity ellipsoid less important than in NGC 3379.) If we accept the Ciardullo *et al.* mass estimate for NGC 3379, then our new result for the mass of NGC 3384 implies that at least a third of the dynamical mass measured by Schneider (1985) has been accounted for in and around the two bright galaxies.

Since the rotation curve defined by the NGC 3384 PNs remains nearly flat to the edge of our sample, it is likely that the total mass associated with this galaxy is substantially greater than the value quoted above. The same might of course be true of NGC 3379, although there is no prima facie reason for believing so; Ciardullo *et al.* (1993) showed that the data do not compel M/L to increase with radius in this galaxy. If we require that half of the additional mass derived by Schneider (1985) be associated with NGC 3384 alone, then the dark halo around this galaxy would have to extend to $\sim 45$ kpc from the galaxy's center. This is comfortably within the limit of $\sim 60$ kpc estimated by Schneider (1991).

## 5. SUMMARY

We have measured radial velocities for 68 PNs around the SB0 galaxy NGC 3384 in the Leo I group. The PN system exhibits strong rotation and a flat rotation curve extending to at least 7 kpc from the galaxy center. We infer a lower limit of $\sim 0.7 \times 10^{11} M_\odot$ for the mass of this galaxy. Given Ciardullo *et al.*'s (1993) estimate of a comparable mass for the nearby elliptical galaxy NGC 3379, we infer that at least a third of the mass of the NGC 3379/3384 group – as determined by Schneider (1985) from the kinematics of the 200 kpc HI ring – has been accounted for in and around the two bright galaxies. Further observations of the PN systems of both NGC 3379 and 3384 are scheduled and should reveal the full extent of the dark matter halos around these galaxies.

The imaging Fabry-Perot is an almost ideal instrument for these observations, allowing one to detect and simultaneously measure the velocity of tens or hundreds of PNs in a few nights of observation. Our map of the rotational velocity field of NGC 3384 extends much farther than rotation curves derived from stellar absorption line spectra (e.g. Franx, Illingworth & Heckman 1989). While some early-type galaxies contain rings of HI gas that can be used to constrain their mass (e.g. van Driel 1987), the resulting rotation curves typically extend over only a short radial range; and many galaxies contain only negligible amounts of HI (see Bregman *et al.* 1992 for the most recent attempt to detect HI in NGC 3384). Thus, PNs provide one of the most promising routes toward mapping dark matter in early-type galaxies.

## ACKNOWLEDGEMENTS

This work was supported by NSF grant AST-9318617 to DM and by a Grant-in-Aid of Research from the National Academy of Sciences through Sigma Xi. The RFP was developed with support from Rutgers University and the National Science Foundation, under grant AST 83-19344. The director and staff of CTIO provided superb levels of technical assistance for the operation and maintenance of the RFP at CTIO, and for these observations.

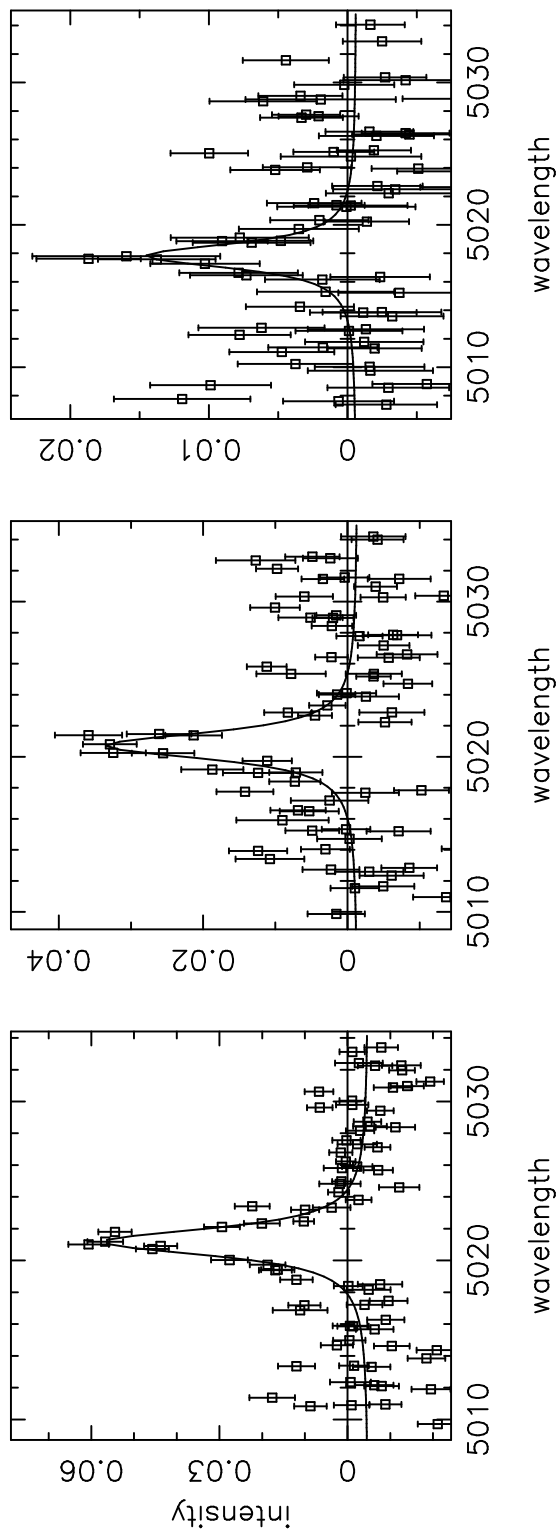

Figure 1. Spectra for three of the planetary nebulae observed in this study. They are, from left to right, nos. 3 ($m_{5007} = 25.7$), 24 (26.1) and 77 (26.7) from the list of Ciardullo *et al.* (1989). The wavelength is measured in angstroms and the intensity is the DAOPHOT flux.

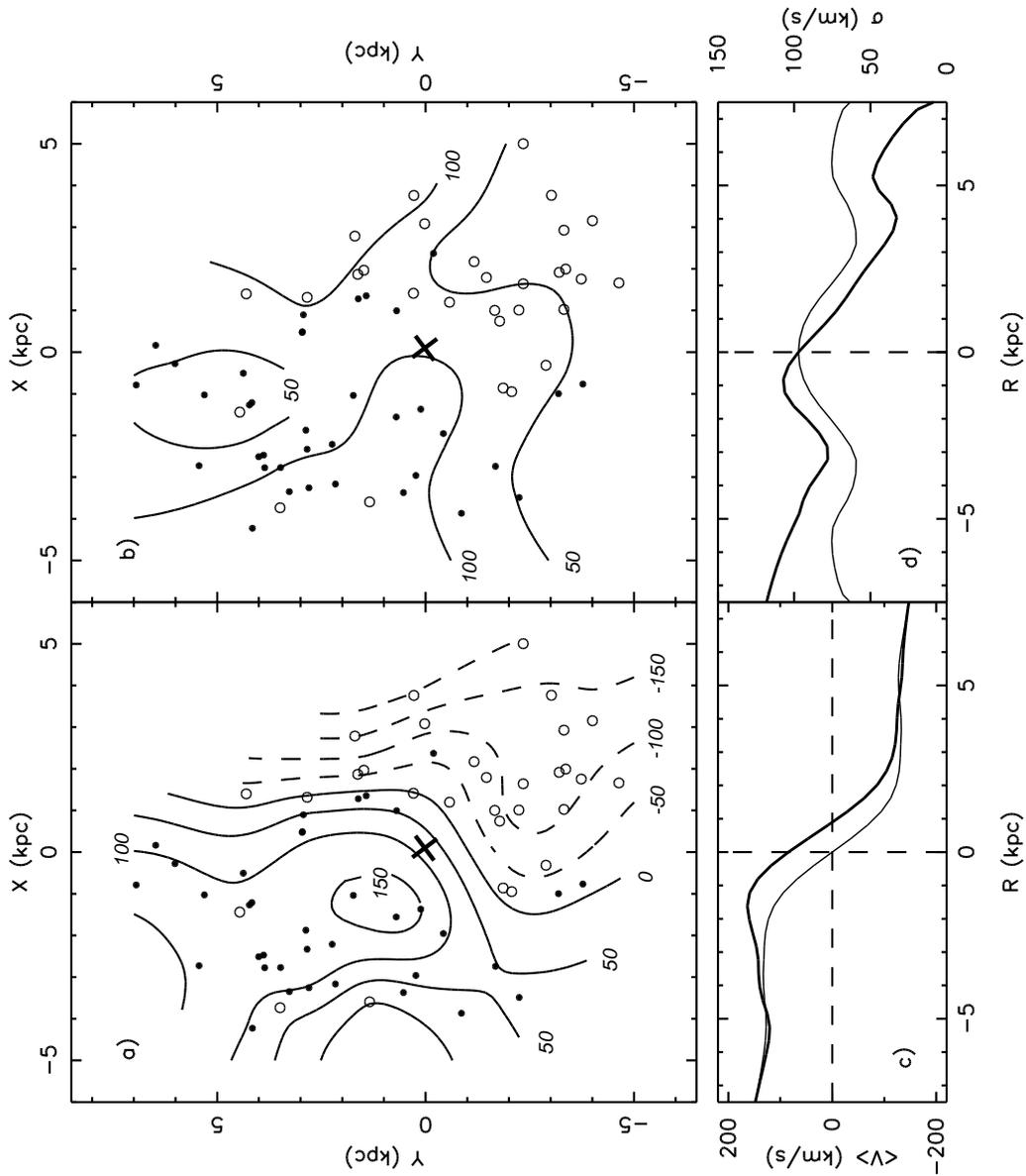

Figure 2. (a) Contours of constant line-of-sight, mean velocity for the NGC 3384 planetary nebula system. East is down and north is to the right. The cross indicates the position and orientation of the galaxy, with the major axis extending from upper left to lower right. Filled/open circles are PNs with positive/negative velocities. (b) Contours of costant line-of-sight velocity dispersion. Symbols have the same meanings as in Fig. 2a. (c) Heavy line: mean line-of-sight velocity along the galaxy major axis; thin line: antisymmetrized profile. (d) Heavy line: line-of-sight velocity dispersion along the galaxy major axis; thinline: symmetrized profile.